\documentclass[aps,preprint,preprintnumbers,amsmath,amssymb,UTF8]{revtex4}
\usepackage{amsmath,mathrsfs,amsbsy,color,graphicx,bm,amsthm,amsfonts}
\usepackage{bbm}
\usepackage{times}
\usepackage{units}
\usepackage{dcolumn}
\usepackage{graphicx}
\usepackage{epsfig}
\usepackage{epstopdf}
\usepackage{booktabs}
\usepackage{hyperref}
\hypersetup{
    colorlinks=true,
    linkcolor=red,
    anchorcolor=green,
    citecolor=blue,
    CJKbookmarks=true
}
\DeclareMathSymbol{\shortminus}{\mathbin}{AMSa}{"39}
\usepackage{mathrsfs}
\usepackage{braket}
\usepackage{amssymb}
\usepackage{txfonts}
\usepackage{float}
\usepackage{enumitem}
\usepackage{multirow}
\usepackage{tabularx}
\usepackage[caption=false]{subfig}

\begin{document}
\title{ Quantum advantage prediction  in turbulent free-space quantum illumination}

\author{Yu Tang}
\thanks{These authors contributed equally to this work.}
 \affiliation{Department of Physics, and Collaborative Innovation Center for Quantum Effects and Applications, Hunan Normal University, Changsha 410081, People’s Republic of China}
 \affiliation{Hunan Research Center of the Basic Discipline for Quantum Effects and Quantum Technologies, Hunan Normal University, Changsha 410081, People’s Republic
of China}

\author{Beining Xia}
\thanks{These authors contributed equally to this work.}
\affiliation{Department of Physics, and Collaborative Innovation Center for Quantum Effects and Applications, Hunan Normal University, Changsha 410081, People’s Republic of China}
\affiliation{College of Information Science and Engineering, Hunan Normal University, Changsha, Hunan 410081, People’s Republic of China}

\author{Qianqian Liu}
\affiliation{Department of Physics, Xiangtan University,
Xiangtan 411105, Hunan Province, China}
       
\author{Cuihong Wen}
\affiliation{Department of Physics, and Collaborative Innovation Center for Quantum Effects and Applications, Hunan Normal University, Changsha 410081, People’s Republic of China}
\affiliation{College of Information Science and Engineering, Hunan Normal University, Changsha, Hunan 410081, People’s Republic of China}

\author{Heng Fan}
\email{hfan@iphy.ac.cn (Corresponding authors)}
\affiliation{Institute of Physics, Chinese Academy of Sciences, Beijing 100190, China}
\affiliation{School of Physical Sciences, University of Chinese Academy of Sciences, Beijing 100049, China}

\author{Jieci Wang}
\email{jcwang@hunnu.edu.cn (Corresponding authors)}
\affiliation{Department of Physics, and Collaborative Innovation Center for Quantum Effects and Applications, Hunan Normal University, Changsha 410081, People’s Republic of China}
 \affiliation{Hunan Research Center of the Basic Discipline for Quantum Effects and Quantum Technologies, Hunan Normal University, Changsha 410081, People’s Republic
of China}

\begin{abstract}

Quantum illumination offers a significant theoretical advantage for target detection in high background noise environments. However, its practical deployment in free-space channels is hindered by atmospheric turbulence. Stochastic fluctuations in atmospheric turbulence inevitably degrade the quantum signature, rendering the real-time evaluation of quantum advantage under such dynamic conditions a critical yet unresolved challenge. To circumvent the reliance on costly direct turbulence measurements, we propose a physics-driven  framework that integrates Kolmogorov-Arnold networks directly bridge macroscopic meteorological observations with microscopic quantum channel dynamics. Trained on 105,120 samples from 12 climatically diverse sites and validated on 26,280 unseen samples from three extreme boundary conditions (arid continental, tropical maritime, high-altitude plateau), our approach establishes a physically consistent mapping from standard meteorological variables to the temporal evolution of the quantum advantage.  This end-to-end system dynamically quantifies the degradation of quantum advantage across diverse turbulence conditions. 
Our results provide a rigorous theoretical and data-driven pathway for environmental adaptation, facilitating the transition of quantum radar networks from proof-of-principle demonstrations to all-weather operational systems.

\end{abstract}
\vspace*{0.5cm}
\maketitle
\section{Introduction}
Quantum illumination (QI) \cite{Lloyd:2008nwf,Tan:2008oho,Barzanjeh:2015zpa,Karsa:2020toy,Zhao:2024njt,Lopaeva:2013dlj} is a quantum-optical sensing technique that exploits quantum sources to improve the detection of a low-reflectivity object  immersed in a bright thermal background.
Unlike quantum key distribution \cite{Grosshans:2001sfp,Grosshans:2003mgc} and teleportation \cite{Bennett:1992tv,Pirandola:2015nwo}, which require entanglement preservation across a transmission channel, QI can retain a performance advantage even when entanglement is destroyed by loss and noise, due to surviving quantum correlations between the retained idler and the returned signal \cite{Assouly:2022rwa,zhuang2023quantum}.  
This unique property confers robustness beyond classical limits, especially within high-loss, low signal-to-noise ratio regimes. 
The resilience of QI positions it as a suitable candidate for next-generation quantum radar and remote sensing systems, enabling operational capabilities that exceed those of classical sensors.

However, the transition of QI from controlled laboratory testbeds to free-space optical communication is significantly affected by a critical physical barrier: the atmospheric turbulence channel \cite{Ursin:2006jya,Scheidl:2009lla,Fedrizzi:2009uiw,Capraro:2012ggc,Yin:2012vzv,Ma:2012bgk,Peuntinger:2014rrq}. 
In operational regimes, from ground-to-satellite communications to long-range radar, optical signals inevitably propagate through turbulent media characterized by stochastic refractive index homogeneity.
Physically, these disturbances are parameterized by the refractive index structure constant $C_n^2$, which induces fading statistics that dictate the probability distribution of the channel transmittance (PDT) \cite{Semenov:2009kas,Vasylyev:2012bnq,Vasylyev:2017rio,Vasylyev:2016xmk, Vasylyev:2018wiw}. 
Seminal studies have established that the performance of quantum channels is highly sensitive to such PDT \cite{Usenko:2012oin,Semenov:2012peh,Zhuang:2017mxi,Chen:2022jud,Hosseinidehaj:2015ffd}. 
Under strong turbulence, the deep fading phenomenon dominates the averaged error probability, degrades the ideal exponential scaling and substantially reducing the achievable quantum advantage \cite{Zhuang:2017mxi,Zhuang:2016jjt}.
Consequently, the future practical deployment of QI is closely tied to the ability to quantify the degradation of its sensing performance within these complex turbulent channels.

Accurately quantifying this sensing degradation requires high-resolution $C_n^2$ profiles, but empirical measurements remain cost-prohibitive and geographically sparse. 
While recent efforts employ deep learning to predict optical turbulence from accessible meteorological variables \cite{cheinet2011use,qing2016estimating,qing2016use,quatresooz2023continuous,hegde2024modeling,wang2016using,su2020adaptive,butterley2017nowcasting,rudiger2018machine,lionis2022optical,lionis2023supervised,campbell2023machine,jellen2020machine,jellen2021machine,bolbasova2021application,jellen2023hybrid,su2021situ,bi2022optical,xu2021optical,pierzyna2023pi}, these purely data-driven models struggle to generalize across complex, unmapped environments. 
This leaves the field with a critical unresolved challenge for evaluating distributed QI: developing robust cross-regional turbulence forecasting that maintains rigorous physical consistency under sparse-data scenarios.

To address this issue, we propose a physics-driven deep learning framework to evaluate the resilience and quantum advantage of QI under the influence of atmospheric turbulence. 
Specifically, we analytically derive the $C_n^2$ from six routine meteorological parameters based on the Monin-Obukhov similarity theory (MOST) \cite{stull2012introduction}. 
Driven by this rigorous physical dataset, we train a hybrid neural network integrating meteorological Kolmogorov-Arnold networks (KAN) predictions \cite{2024KAN}. 
By encompassing a statistically representative parameter space across diverse climatic regions and extreme meteorological boundary conditions, our dataset formulation enables the framework to capture the complex nonlinear mappings between macroscopic meteorological drivers and optical turbulence, thereby achieving robust cross-regional generalization.
Finally, the deep learning predicted $C_n^2$ profiles are directly embedded in the QI channel model to dynamically quantify the degradation of quantum advantage, thus establishing a direct analytical link from macro-meteorology to physical limits of quantum sensing. 
This work establishes a quantitative mapping from meteorological fields to QI performance via turbulence statistics.

\section{Physics-driven framework for optical turbulence prediction}\label{ph}
We begin by establishing the framework capable of mapping macroscopic meteorological variables to optical turbulence $C_n^2$. Although future distributed QI deployments inevitably involve complex slant paths, the continuous height-dependent turbulence profiles complicate the derivation of the overall PDT. 
In contrast, a near-ground horizontal channel allows the assumption of a constant $C_n^2$ along the propagation path, defining a uniform PDT while strictly satisfying the boundary conditions of the MOST \cite{stull2012introduction}.
Therefore, as a foundational step, we focus on horizontal free-space channels to simplify the theoretical analysis.
To map meteorological observables into optical turbulence dynamics, we employ the MOST to govern the micrometeorology of the atmospheric surface layer (SL). By assuming constant vertical fluxes, MOST yields the characteristic scaling parameters that drive SL turbulence. 
For future QI systems operating at visible and near-infrared wavelengths \cite{Tan:2008oho,Lopaeva:2013dlj}, refractive index fluctuations are overwhelmingly dominated by thermal variations \cite{hill1980refractive}. Consequently, the overall turbulence strength $C_n^2$ is directly determined by the temperature structure parameter $C_T^2$ through the Gladstone-Dale relation \cite{roddier1981v,andrews2005laser}:
\begin{equation}
C_n^2=\left(\frac{79\times10^{-6}P}{T^2}\right)^2C_T^2,\label{cn2}
\end{equation}
where $P$ and $T$ denote the local atmospheric pressure and temperature, respectively.

To dynamically evaluate these horizontal channels across diverse environments, we utilize the EnergyPlus Weather (EPW) database \cite{epw_database}. 
By providing standardized, hourly data based on Typical Meteorological Years, the EPW dataset excludes anomalous weather events to ensure high statistical reliability.
This globally consistent meteorological baseline establishes a rigorous physical foundation for our MOST-based $C_n^2$ derivations and the continuous tracking of quantum channel dynamics.
The meteorological parameters extracted from the EPW dataset are shown in Table \ref{Tab1}.
\begin{table}[htbp]
\centering
\caption{Meteorological parameters extracted from the EPW}
\label{Tab1}
\renewcommand{\arraystretch}{1.2}
\begin{tabular*}{\linewidth}{@{\extracolsep{\fill}} ccc @{}}
\toprule
Parameter &  Unit  & Description \\
\midrule
$T$ &  $^\circ$C &  Dry Bulb Temperature \\
$T_{dp}$ & $^\circ$C &  Dew Point Temperature \\
$P$ & Pa &  Atmospheric Station Pressure \\
$u$ & $m/s$ &  Wind Speed \\
$L_{\downarrow}$ & $W/m^2$ &  Horizontal Infrared Radiation Intensity \\
$GHI$ & $W/m^2$ &  Global Horizontal Irradiance \\
\bottomrule
\end{tabular*}
\end{table}
Based on the  MOST theory, we utilize the six meteorological parameters from Table \ref{Tab1} to compute $C_n^2$, thus establishing a physics-driven meteorological data-driven framework for the prediction of optical turbulence $C_n^2$ (please see the Supplementary Materials for derivations).

\section{DEEP LEARNING FOR SPATIAL GENERALIZATION ACROSS DIVERSE CLIMATES
}\label{sp}
To establish a statistically representative parameter space for atmospheric dynamics, we train on 105,120 samples across 12 climatically diverse sites (Beijing, Chengdu, Guangzhou, Harbin, Hangzhou, Kunming, Xiamen, Shanghai, Urumqi, Wuhan, Xi’an, and Changsha). 
Cross-regional generalization is rigorously validated using 26,280 unseen samples under three extreme meteorological boundary conditions: arid continental (Hohhot), tropical maritime (Sanya), and high-altitude plateau (Lhasa).
To map the multivariable “meteorology–$C_n^2$" physical relationship, we adopt the KAN model. 
By transferring nonlinear expressivity from fixed node activations to learnable edge weight functions, KAN significantly surpasses traditional multilayer perceptrons in approximation efficiency and physical interpretability. 
Quantitative evaluation ($R_{xy}$, MSE, MAE, and bias) confirms that KAN achieves superior fitting accuracy with a highly concentrated error distribution. 
Crucially, time-series analysis demonstrates KAN's heightened sensitivity to abrupt $C_n^2$ fluctuations, effectively mitigating the systematic underestimation and temporal lag characteristic of intense turbulence variations (please see the Supplementary Materials  for experimental results).

\section{Degradation of QI Advantage in Atmospheric Turbulence}\label{degra}
\subsection{QI protocol under atmospheric channels}\label{qipro}
In this section,  we integrate an atmospheric attenuation model into the quantum illumination protocol and evaluate the impact of channel losses. We then extend the error probability analysis to the turbulence regime, analytically showing how turbulence-induced fading degrades the quantum advantage bounds. 
We employ the two-mode squeezed vacuum (TMSV) state, characterized by a mean signal photon number $N_S$, as the bipartite continuous-variable quantum probe. 
In the standard QI protocol, the signal mode $\hat{a}_S$ interrogates a region containing a bright thermal bath $N_B \gg 1$, while the idler $\hat{a}_I$ is retained for joint measurement. 
Under the target-absent hypothesis $H_0$, the return mode comprises solely the thermal background $\hat{a}_R = \hat{a}_B$. 
Under the target-present hypothesis $H_1$, the target weakly reflects $\kappa \ll 1$, and the return mode is modeled via an effective beam-splitter interaction with the environment
\begin{equation}
\hat{a}_R=\sqrt{\kappa}\hat{a}_S+\sqrt{1-\kappa}\hat{a}_B.\label{hata}
\end{equation}

The overall transmissivity and reflectivity $\kappa$ can be estimated using the radar
equation.
This equation expresses the power $P_R$ of the return signal in terms of the signal power $P_T$ at the transmitter, the cross section $\sigma$ of the target, the range $L$ of the target, the gain of the transmit antenna $G$, the receiving antenna collection area $A_R$, and the form factor $F$, which describes the transmissivity of the space between the radar and the target
\cite{skolnik1980introduction}:
\begin{equation}
P_R=\frac{GF^4A_R\sigma}{(4\pi)^2L^4}P_T.\label{kk}
\end{equation}
It is clear that $\kappa$ also provides the ratio between received and transmitted power, so that Eq. (\ref{kk}) leads to 
\begin{equation}
\kappa=\frac{P_R}{P_T}=\frac{GF^4A_R\sigma}{(4\pi)^2L^4}.\label{kkk}
\end{equation}

Here we consider the atmospheric attenuation channel model, where $C_n^2$ remains uniform along the horizontal free-space path. 
Let $\tau$ denote the single-trip atmospheric transmittance. 
Since the signal undergoes a round-trip propagation, the overall atmospheric two-way transmittance is $\tau^2$, which yields the relation $F^4 = \tau^2$.
Since our primary focus is the atmospheric attenuation mechanism, the specific hardware and geometric factors of realistic radar systems are abstracted into a static baseline constant $\kappa_0 = G A_R \sigma/{(4\pi)^2 L^4} \ll 1$.
Consequently, the dynamic overall reflectivity is governed entirely by the atmospheric transmittance fluctuation, yielding $\kappa = \kappa_0 \tau^2$.

\subsection{QI advantage in atmospheric turbulence}\label{qiad}
The typical operational regime of QI is characterized by low target reflectivity $\kappa \ll 1$, high thermal background noise $N_B \gg 1$, and a low signal photon number per mode $N_S \ll 1$. 
Substituting $\kappa = \kappa_0 \tau^2$ into the model, we obtain the error probability for a protocol employing a TMSV state satisfies the upper bound \cite{Tan:2008oho}:
\begin{equation}
P_{\mathrm{err}}^{\mathrm{QI}}\leqslant \frac{1}{2} e^{-\Lambda \tau^2 N_S/N_B}.\label{pqi}
\end{equation}
Here, $\Lambda = M \kappa_0$ defines the effective mode number, a lumped parameter wherein the massive time-bandwidth product $M$ inherent to QI offsets the extreme baseline geometric loss $\kappa_0$.
This upper limit, derived via the quantum Bhattacharyya bound, is exponentially tight in the limit of a large number of modes $M$.
Notably, the error probability exponent of this bound exhibits a 6 dB advantage over the corresponding limit for a coherent state transmitter operating under identical conditions, yielding \cite{Tan:2008oho}
\begin{equation}
P_{\mathrm{err}}^{\mathrm{CS}}\leqslant \frac{1}{2} e^{-\Lambda \tau^2 N_{S}/4N_{B}}.\label{pcs}
\end{equation}
In the regime of turbulence, the error probability must be averaged across all channel realizations
\begin{equation}
\bar{P}_{\mathrm{err}}=\int_{0}^{\tau_{\mathrm{max}}}d\tau \mathcal{P}(\tau)P_{\mathrm{err}}( \tau).\label{pk}
\end{equation}
Note that $\bar{P}_{\mathrm{err}}$ directly depends on the PDT $\mathcal{P}(\tau)$, and therefore on the properties of the atmospheric channel.
Specifically, the integral in Eq. (\ref{pk}) can be dominated by the near-zero transmittance tail of the PDT under strong turbulence (characterized by large $C_n^2$ values), a phenomenon known as deep fading. 
This deep fading destroys the ideal exponential decay of the error probability, causing the 6 dB error probability exponent advantage of QI to collapse \cite{Zhuang:2017mxi,Zhuang:2016jjt}. 
Therefore, the QI advantage must be quantified by directly comparing the absolute averaged error probability, $\mathcal{R}=\bar{P}_\mathrm{err}^\mathrm{CS}/\bar{P}_\mathrm{err}^\mathrm{QI}$.
In this paper, we work with the PDT of the elliptic beam model \cite{Vasylyev:2016xmk}. 
This model takes into account the deflection and deformation of a Gaussian beam caused by turbulence in atmospheric channels and shows good agreement with experimental free-space channels \cite{Vasylyev:2017rio}.
The implementation of the model for calculating the error probability $\bar{P}_{\mathrm{err}}$ is presented in the Supplementary Materials, together with further QI channel parameters.
\begin{figure}[!htbp]
\centering
\includegraphics[width=0.9\columnwidth]{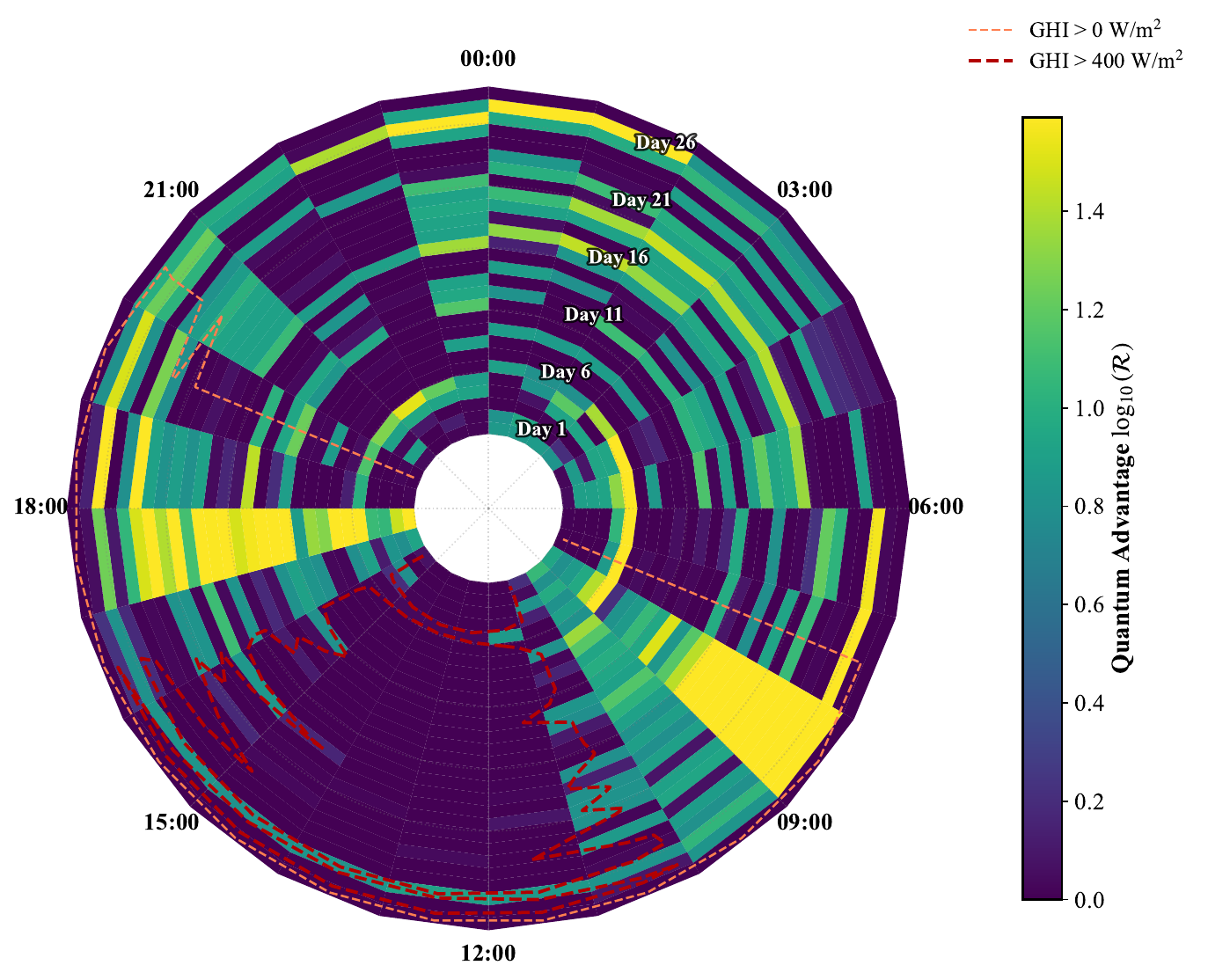}
\caption{\textbf{Diurnal phase evolution of QI advantage under meteorological forcing.} The polar clock maps the degradation of QI channels across days and local hours. 
The logarithmic QI advantage is truncated at the 95th percentile to highlight steady states. 
Dashed contours mark GHI thresholds: twilight transition $\mathrm{GHI} > 0\,\mathrm{W/m^2}$ and strong daytime convection $\mathrm{GHI} > 400\,\mathrm{W/m^2}$.}
\label{clock}
\end{figure}\noindent
\begin{figure}[!htbp]
    \centering
    \includegraphics[width=0.45\textwidth]{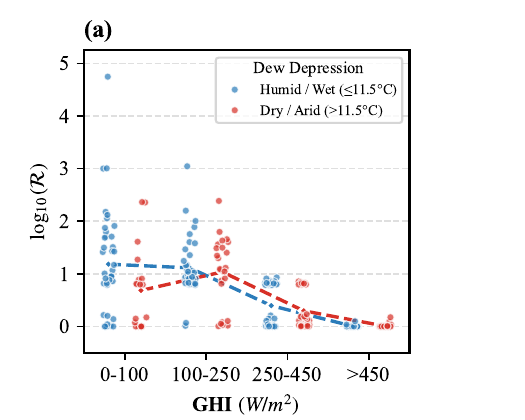}
    \includegraphics[width=0.45\textwidth]{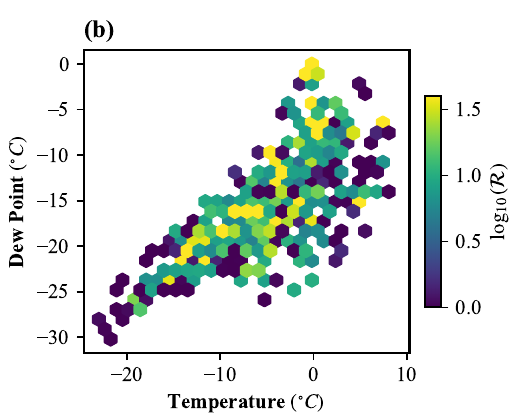} 
    \caption{\textbf{Scatter and density distributions of QI advantage under varying meteorological parameters. (a)}  The $\log_{10}\mathcal{R}$ evaluated against GHI. 
    The data are partitioned based on the median dew depression to present the joint distribution of $\log_{10}\mathcal{R}$ under different solar radiation and atmospheric moisture conditions. Dashed lines connect the mean values of $\log_{10}\mathcal{R}$ for each category. 
    \textbf{(b)} A two-dimensional hexbin map illustrating the 95th percentile of $\log_{10}\mathcal{R}$ across the thermodynamic phase space defined by dry bulb temperature and dew point.}
    \label{fig2}
\end{figure}
\subsection{Meteorological-driven dynamic degradation of QI advantage}
In this section, we bridge macroscopic meteorological phenomena with QI target-detection performance through the $C_n^2$ profiles predicted by our KAN architecture. 
By utilizing these predictions to parameterize the PDT of the atmospheric channel, we systematically investigate how diverse meteorological variables modulate optical turbulence and drive the dynamic degradation of QI advantage. 
To illustrate this interplay, we employ the February 2001 meteorological dataset from the Hohhot station, allocated as part of our test set, to explicitly map the coupled fluctuations between specific meteorological drivers and the resultant erosion of quantum performance.

As illustrated in Fig. \ref{clock}, the QI advantage is projected onto a polar grid, revealing a strong diurnal modulation by global horizontal irradiance (GHI). 
When $\mathrm{GHI}$ exceeds $400\,\mathrm{W/m^2}$ during the strong convective daytime, intense solar heating drives turbulent updrafts, collapsing the QI advantage to near-classical baselines. 
Conversely, when $\mathrm{GHI}$ falls to $0\,\mathrm{W/m^2}$ at nighttime, radiative cooling suppresses thermal turbulence, allowing a moderate recovery of the quantum signal, though residual mechanical turbulence persists.	
Critically, a “golden window" emerges during twilight transitions $0 < \mathrm{GHI} \le 100\,\mathrm{W/m^2}$. 
In this regime, near-zero net surface heating minimizes the surface-air temperature gradient, while wind speeds are typically weak; the simultaneous suppression of both thermal and mechanical turbulence drives the quantum advantage to its diurnal maximum.
\begin{figure}[!htbp]
\centering
\includegraphics[width=0.85\columnwidth]{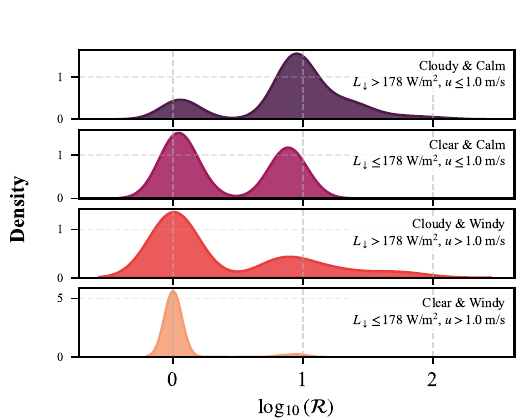}
\caption{\textbf{Probability density distributions of $\log_{10}\mathcal{R}$ in the nocturnal boundary layer ($\mathrm{GHI} = 0\,\mathrm{W/m^2}$).} The dataset is partitioned into four distinct quadrants based on downward longwave irradiance (threshold $L_{\downarrow}=178\,\mathrm{W/m^2}$) and wind speed (threshold $u=1.0\,\mathrm{m/s}$).}
\label{fig3}
\end{figure}\noindent

Although GHI establishes the macroscopic diurnal envelope of the QI advantage, this primary radiative forcing does not act in isolation. 
The optical turbulence emerges from the complex coupling of multidimensional meteorological parameters.
Focusing exclusively on daytime conditions $\mathrm{GHI} > 0\,\mathrm{W/m^2}$, the degradative impact of GHI on $\log_{10}\mathcal{R}$ is nonlinearly modulated by dew depression $T-T_{dp}$ via surface energy partitioning (Fig. \ref{fig2}a).
Under low-to-moderate radiation $0–250\,\mathrm{W/m^2}$, the humid regime maintains a stable QI advantage, as solar energy is predominantly dissipated through evaporation, effectively limiting surface temperature rise and suppressing strong thermal gradients. 
Conversely, the arid regime exhibits a transient spike in $\log_{10}\mathcal{R}$; rapid surface heating drives the atmospheric surface layer toward a near-neutral condition (where the potential temperature gradient approaches zero), momentarily minimizing $C_n^2$.
However, extreme solar forcing $\mathrm{GHI} > 450\,\mathrm{W/m^2}$ induces excessive optical turbulence that indiscriminately collapses the QI advantage toward the classical limit across both regimes.
Fig. \ref{fig2}b further constrains this coupling within a thermodynamic phase space. 
The peak $\log_{10}\mathcal{R}$ values (95th percentile) tightly localize to a narrow characteristic band, providing direct evidence for the coupled modulation of QI by ambient temperature and moisture.

While the preceding analysis elucidates the thermodynamic coupling under daytime solar forcing $\mathrm{GHI} > 0\,\mathrm{W/m^2}$, the focus now shifts to the nocturnal boundary layer $\mathrm{GHI} = 0\,\mathrm{W/m^2}$. 
In the absence of solar radiation, the SL optical turbulence is primarily modulated by horizontal infrared radiation intensity $L_{\downarrow}$ (downward longwave radiation from the entire atmosphere) and mechanical wind shear $u$.
As shown in Fig. \ref{fig3}, the nocturnal QI advantage is influenced by the competition between $L_{\downarrow}$ and $u$. 
Under cloudy and calm conditions, downward longwave radiation buffering limits surface cooling. 
This dual absence of steep thermal gradients and mechanical mixing globally minimizes $C_n^2$, driving the probability density of $\log_{10}\mathcal{R}$ to its absolute peak.
Conversely, under clear and calm conditions, uninhibited surface radiative cooling constructs a strong thermal inversion.
While the QI advantage survives without wind shear, this steep vertical temperature gradient persists in the absence of wind shear, this steep vertical temperature gradient constitutes an exceptionally fragile physical state.
\begin{figure}[!htbp]
\centering
\includegraphics[width=0.85\columnwidth]{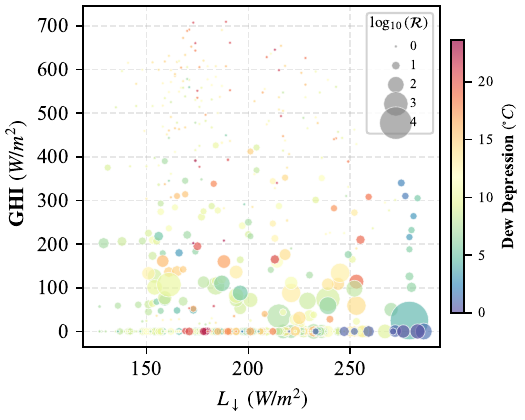}
\caption{\textbf{Multidimensional phase space representation of the QI advantage.} 
The magnitude of $\log_{10}\mathcal{R}$ is projected onto the orthogonal axes of shortwave irradiance $\mathrm{GHI}$ and downward longwave irradiance $L_{\downarrow}$. 
The background thermodynamic state is parameterized by the dew depression. }
\label{fig4}
\end{figure}\noindent
Moreover, the decisive collapse of the nocturnal channel is primarily driven by mechanical forcing.
In the cloudy and windy regimes, wind shear $u > 1.0\,\mathrm{m/s}$ induces forced mixing, but the weak temperature gradient restrains $C_n^2$ amplification, preserving a broad distribution buffer. 
In stark contrast, when mechanical shear acts upon the strong thermal inversion $L_{\downarrow} \le 178\,\mathrm{W/m^2}$, it triggers shear-induced instability. 
This kinetic shear violently disrupts the stratified layers, driving massive temperature fluctuations. 
The resulting explosion of $C_n^2$ entirely crushes the probability distribution toward the classical limit $\log_{10}\mathcal{R} \to 0$. 
These dynamics rigorously demonstrate that the disruption of strong thermal inversions by mechanical shear acts as a mechanism for channel collapse in nocturnal QI.

Finally, we extend our analysis to a higher-dimensional phase space, mapping the QI advantage across the orthogonal axes of shortwave $\mathrm{GHI}$ and longwave $L_{\downarrow}$ radiative forcings, under the strict constraint of thermodynamic moisture.
As illustrated in Fig. \ref{fig4}, the global maxima of the QI advantage $\log_{10}\mathcal{R}$ strictly demand a precise energetic and thermodynamic state: zero shortwave irradiance $\mathrm{GHI} = 0\,\mathrm{W/m^2}$, strong longwave buffering $L_{\downarrow} > 250\,\mathrm{W/m^2}$, and minimal dew depression $T-T_{dp} \to 0$. 
This localized condition effectively freezes optical turbulence. 
Conversely, excessive shortwave forcing introduces strong thermal excitation that unconditionally shatters this equilibrium, crushing the channel toward the classical limit.

\section{Discussion}\label{dis}
In this work, we have established a deterministic analytical mapping from accessible macroscopic meteorological drivers directly to the microscopic quantum channel dynamics of free-space QI.
By anchoring a physics-driven deep learning framework within the rigorous boundary conditions of MOST, we bypassed the prohibitive reliance on empirical $C_n^2$ measurements. 
This approach enables the real-time assessment of turbulence-induced deep fading, revealing the fundamental degradation mechanisms of the QI advantage under atmospheric turbulence.
Consequently, quantifying and preserving the QI advantage in free-space configurations is fundamentally a coupled macroscopic-microscopic atmospheric physics challenge.
Ultimately, this physics-driven mapping enables the continuous quantitative evaluation of the sensing performance under simulated atmospheric fluctuations.
By parameterizing the boundary limits of channel viability without requiring extensive field campaigns, this framework provides a reproducible theoretical benchmark and a necessary analytical foundation for the long-term architectural design of future free-space quantum sensing configurations.

\acknowledgments
This work was supported by the National Natural Science
Foundation of China (Grants No. 12374408, No. 12475051,  and No. 12547147); the science and technology innovation Program of Hunan Province under grant No. 2024RC1050; the Natural Science Fund of Hunan Province under Grant No. 2026JJ20019; and the China Postdoctoral Science Foundation (Grant No. 2025M783393).

\section{Appendix}
\subsection{From meteorological parameters to $C_n^2$}
In this section, we detail the analytical derivation of the optical turbulence profile $C_n^2$ based on the Monin-Obukhov similarity theory (MOST). 
Specifically, we construct this physics-driven mapping using exactly six standard macroscopic meteorological observables: dry bulb temperature $T$ ($^\circ$C), dew point temperature $T_{dp}$ ($^\circ$C), atmospheric pressure $P$ (Pa), wind speed $u$ ($m/s$), global horizontal irradiance $\mathrm{GHI}$ ($W/m^2$), and horizontal infrared radiation $L_{\downarrow}$ ($W/m^2$).

\subsubsection{ Derivation of thermodynamic state variables}
The initiation of the mathematical framework requires the translation of the observed, bulk meteorological parameters into the fundamental thermodynamic state variables that govern energy exchange, buoyancy, and density within the atmospheric surface layer.

The observed atmospheric pressure $P$ is routinely recorded in diverse units depending on the instrumentation,
for strict consistency with the subsequent energy balance and psychrometric equations, the pressure must be standardized into kilopascals $P_{kPa} = P/1000$.
The moisture content of the atmosphere is a critical driver of latent heat flux and atmospheric stability. 
The actual vapor pressure of the air, $e_a$ (kPa), is determined from the measured temperature of the dew point $T_{dp}$ \cite{allen1998crop}
\begin{equation}
e_a=0.6108\cdot\exp\left(\frac{17.27\cdot T_{dp}}{T_{dp}+237.3}\right).\label{ea}
\end{equation}
Similarly, the saturation vapor pressure, $e_s$ (kPa), which represents the maximum moisture-holding capacity of the air at a given temperature, is calculated using the dry bulb temperature $T$ \cite{allen1998crop}
\begin{equation}
e_s=0.6108\cdot\exp\left(\frac{17.27\cdot T}{T+237.3}\right).\label{es}
\end{equation}
The difference between these two values, $e_s - e_a$, is the vapor pressure deficit (VPD). 
The VPD is the primary thermodynamic driving force for water evaporation and latent heat flux, representing the atmosphere's aerodynamic drying power.
To accurately assess the buoyancy of the air parcel, which is required to determine atmospheric stability and the generation of turbulent kinetic energy, the specific humidity and the virtual temperature must be derived. 
Specific humidity $q$ is expressed as \cite{brutsaert2013evaporation}
\begin{equation}
q=\frac{0.622\cdot e_a}{P_{kPa}-0.378\cdot e_a}. \label{shq}
\end{equation}
Because water vapor is less dense than dry air, a moist air parcel will experience greater buoyancy than a dry air parcel at the exact same temperature and pressure. To utilize the ideal gas law for moist air without altering the gas constant, we need to use the virtual temperature $T_v$ \cite{brutsaert2013evaporation},
\begin{equation}
T_v=(T+273.15)\cdot(1+0.61\cdot q).\label{vt}
\end{equation}
The accurate determination of air density is paramount, as it directly scales both the momentum flux and the sensible heat flux in the MOST
\begin{equation}
\rho=\frac{P}{R_d\cdot T_v},\label{rho}
\end{equation}
where $R_d$ is the specific gas constant for dry air, giving a value of $\unit[287.05]{\mathrm{J}\cdot\mathrm{kg}^{-1}\cdot\mathrm{K}^{-1}}$.
\subsubsection{Surface energy balance closure}
The generation of atmospheric turbulence in the boundary layer is driven by the partitioning of available energy at the Earth's surface. 
The second step of the derivation defines this surface energy budget, utilizing the provided GHI and downward longwave radiation $L_{\downarrow}$.
The Earth's surface continuously emits longwave radiation into the atmosphere. 
According to the Stefan-Boltzmann law, the upward longwave radiation $L_{\uparrow}$ is proportional to the fourth power of the absolute surface temperature. 
However, as the EPW files in question do not include surface skin temperature data relating to aerodynamics, our framework approximates surface emission using dry bulb temperature $T$ and effective surface emissivity ($\epsilon_s=0.98$)
\begin{equation}
L_\uparrow=\epsilon_s\cdot\sigma_{s}\cdot(T+273.15)^4, \label{lup}
\end{equation}
where $\sigma_s$ is the Stefan-Boltzmann constant.

Net radiation, $R_n$ ($W/m^2$), quantifies the total radiative energy available at the surface. 
It is the sum of the net shortwave radiation and the net longwave radiation. 
The net shortwave component is simply the measured GHI adjusted for the surface albedo ($\alpha=0.23$) \cite{allen1998crop}, representing the fraction of reflected solar energy. 
The complete net radiation equation is given by
\begin{equation}
R_n=\begin{pmatrix}1-\alpha
\end{pmatrix}\cdot GHI-
\begin{pmatrix}L_\uparrow-L_\downarrow
\end{pmatrix}.\label{rn}
\end{equation}
However, not all net radiation is available to be partitioned into turbulent fluxes; a fraction of the energy is absorbed downward into the substrate as soil heat flux $G_s$ ($W/m^2$).
Because continuous measurement of $G_s$ requires buried heat flux plates, meteorological models typically parameterize $G_s$ as a fraction of the net radiation \cite{allen1998crop}
\begin{equation}
G_s=
\begin{cases}
0.1\cdot R_n & \mathrm{for}\quad\mathrm{}R_n>0\mathrm{,daytime} \\
0.5\cdot R_n & \mathrm{for}\quad\mathrm{}R_n<0\mathrm{,nighttime}
\end{cases}.\label{dt}
\end{equation}
With $R_n$ and $G_s$ defined, the total available energy at the surface, $A$ ($W/m^2$), is simply the residual
\begin{equation}
A=R_n-G_s.\label{ava}
\end{equation}
This available energy subsequently be partitioned into the latent heat flux $LE$ and the sensible heat flux $H$.
\subsubsection{The Advection-Aridity evapotranspiration model}
To determine the sensible heat flux, which directly scales the temperature fluctuations causing optical turbulence, one must first solve for the latent heat flux. 
In humid, well-watered environments, latent heat consumes the vast majority of available energy. 
In arid environments, sensible heat dominates. 
To accurately estimate actual evapotranspiration without the aid of soil moisture sensors or stomatal conductance data, we employ the Advection-Aridity (AA) model \cite{brutsaert1979advection}.

The AA model is founded on Bouchet's complementary hypothesis. 
As the landscape dries, actual evapotranspiration $LE$ decreases below the wet-environment evaporation $LE_{PT}$.
The surplus energy is converted to sensible heat, warming and drying the boundary layer, which in turn increases the atmospheric evaporative demand as represented by the Penman evaporation $LE_{Penman}$.
The complementary relationship between $LE$, $LE_{PT}$, and $LE_{Penman}$ then allows estimation of actual evapotranspiration $LE$ as \cite{brutsaert1979advection}
\begin{equation}
LE=2\cdot LE_{PT}-LE_{Penman}.\label{acle}
\end{equation}

The potential evapotranspiration of a wet environment, where aerodynamic forcing is minimal and radiation dominates, is calculated using the Priestley-Taylor equation \cite{priestley1972assessment}
\begin{equation}
LE_{PT}=1.26\cdot\frac{\Delta}{\Delta+\gamma}\cdot A. \label{lpt}
\end{equation}
Here, $\Delta$ is the slope of the saturation vapor pressure curve  \cite{allen1998crop}
\begin{equation}
\Delta=\frac{4098\cdot e_s}{(T+237.3)^2},\label{edlta}
\end{equation}
and $\gamma$ is the psychrometric constant
\begin{equation}
\gamma=\frac{C_p\cdot P}{0.622\cdot L_v},\label{gamma}
\end{equation}
where $C_p=\unit[1013]{\mathrm{J}\cdot\mathrm{kg}^{-1}\cdot\mathrm{K}^{-1}}$ is the specific heat of moist air, and $L_v$ is the latent heat of vaporization of water, which exhibits a slight linear dependence on air temperature \cite{allen1998crop}
\begin{equation}
L_v=(2.501-0.002361\cdot T)\times10^6.\label{latnet}
\end{equation}

The apparent potential evaporation, representing the atmospheric drying demand, is calculated using the Penman combination equation in its impedance form.
 This equation partitions the available energy into a radiative term and an aerodynamic term \cite{penman1948natural}
 \begin{equation}
 LE_{Penman}=\frac{\Delta}{\Delta+\gamma}\cdot A+\frac{\gamma}{\Delta+\gamma}\cdot E_A,\label{penman}
 \end{equation}
where $E_A$ represents the aerodynamic drying power of the atmosphere 
\begin{equation}
E_A=\frac{\rho\cdot C_p}{\gamma\cdot r_a}\cdot
\begin{pmatrix}e_s-e_a
\end{pmatrix}.\label{ea}
\end{equation}
The aerodynamic resistance $r_a$ is estimated using the standard logarithmic wind profile under the assumption of neutral atmospheric stability \cite{allen1998crop}
\begin{equation}
r_a=\frac{\ln(z_u/z_{0m})\cdot\ln(z_t/z_{0h})}{k^2\cdot u},\label{rara}
\end{equation}
where $z_u$ and $z_t$ are the measurement heights for wind and temperature, $z_{0m}$ and $z_{0h}$ are the aerodynamic roughness lengths for momentum and heat, and $k=0.4$ is the von Kármán constant.
The measurement heights are specified as $z_u=\unit[10]{\mathrm{m}}$ and $z_t=\unit[1.5]{\mathrm{m}}$, in accordance with World Meteorological Organization (WMO) and Chinese operational standards. 
The roughness lengths are assigned as $z_{0m}=\unit[0.03]{\mathrm{m}}$, the FAO-56 \cite{allen1998crop} recommended value for short grass, and $z_{0h}=\unit[0.003]{\mathrm{m}}$ , adopting the common approximation $z_{0m}/z_{0h}=10$  to reflect the less efficient turbulent transport of heat compared to momentum.
This parameterisation is a practical simplification widely accepted in operational evaporation models.

With the actual latent heat flux $LE$ determined via the complementary AA model, the sensible heat flux $H$ is extracted as the residual of the surface energy balance
\begin{equation}
H=A-LE.\label{hf}
\end{equation}
Finally, the mass flux of water vapor, $E$ ($\mathrm{kg}\cdot \mathrm{m}^{-2}\cdot \mathrm{s}^{-1}$), is isolated by dividing the latent energy flux by the latent heat of vaporization
\begin{equation}
E=\frac{LE}{L_v}.
\end{equation}

\subsubsection{MOST iterative resolution}
Having derived the bulk sensible heat flux $H$ and mass flux $E$, we now transition from macro-scale thermodynamics to the micro-scale dynamics of the atmospheric boundary layer. 
This transition is mediated by MOST \cite{stull2012introduction}, which posits that within the surface layer, dimensionless turbulence statistics—such as velocity and temperature gradients—are universal functions of a single dimensionless stability parameter $\zeta = z/L_{\mathrm{Ob}}$, where $z$ is the height and $L_{\mathrm{Ob}}$ is the Obukhov length.
The fundamental scaling parameters of MOST are the friction velocity $u_*$, which characterizes the mechanical shear stress at the surface, and the Obukhov length $L_{\mathrm{Ob}}$, which characterizes the relative dominance of mechanical shear versus thermal buoyancy.

The calculation of $u_*$ and $L_{\mathrm{Ob}}$ presents a classic mathematical closure problem. 
The friction velocity $u_*$ relies on the universal stability function for momentum $\Psi_m(\zeta)$, which requires knowledge of $\zeta$. 
However, the Obukhov length $L_{\mathrm{Ob}}$ itself is a function of the cube of the friction velocity $u_*^3$ \cite{stull2012introduction}. 
Because $u_*$ depends on $L_{\mathrm{Ob}}$, and $L_{\mathrm{Ob}}$ depends on $u_*$, the system of equations is highly non-linear and transcendental. 
It cannot be solved via a direct analytical formula. 
Consequently, an iterative numerical algorithm must be employed.

The iterative loop is initialized by assuming an entirely neutral atmosphere. 
In a neutral boundary layer, thermal buoyancy is zero, mechanical shear dominates completely, and the Obukhov length approaches infinity $L_{\mathrm{Ob}} \to \infty$. 
Consequently, the stability parameter $\zeta \to 0$, and the momentum stability correction function evaluates to zero $\Psi_m(0) = 0$.
The initial guess for friction velocity $u_{*0}$ is therefore derived solely from the standard logarithmic wind profile without stability corrections \cite{stull2012introduction}
\begin{equation}
u_{*0}=\frac{k\cdot u}{\ln(z_u/z_{0m})}.\label{frau}
\end{equation}
The classical formulation of $L_{\mathrm{Ob}}$ uses the flux of virtual potential temperature. 
The mathematical framework translates this into terms of sensible heat $H$ and water vapor mass flux $E$ \cite{stull2012introduction}
\begin{equation}
L_{\mathrm{Ob}}=-\frac{\rho\cdot C_p\cdot T_v\cdot u_*^3}{k\cdot g\cdot[H+0.61\cdot(T+273.15)\cdot C_p\cdot E]}, \label{lob}
\end{equation}
where $g$ is gravitational acceleration.
A negative $L_{\mathrm{Ob}}$ signifies an unstable, convective atmosphere, while a positive $L_{\mathrm{Ob}}$ signifies a stable, stratified atmosphere.
Once $L_{\mathrm{Ob}}$ has been calculated, the dimensionless parameter $\zeta$ can be determined for the ratio $z/L_{\mathrm{Ob}}$.

Based on the sign and magnitude of $\zeta$, the universal stability correction function for momentum $\Psi_m(\zeta)$ is evaluated using empirical forms derived from the 1968 Kansas experiment \cite{businger1971flux},  with subsequent refinements from other field campaigns including Wangara \cite{clarke1971wangara}.

In a stable nocturnal boundary layer $\zeta > 0$, negative buoyancy aggressively suppresses vertical turbulent mixing, compressing the depth of the surface layer. 
The analytical integration is given by \cite{webb1970profile,andreas1988estimating}
\begin{equation}
\Psi_m(\zeta)=-7\cdot\zeta.\label{stab}
\end{equation}
For unstable conditions $\zeta < 0$, the integrated form is \cite{paulson1970mathematical}
\begin{equation}
\Psi_m(\zeta)=2\ln\left(\frac{1+x}{2}\right)+\ln\left(\frac{1+x^2}{2}\right)-2\arctan(x)+\frac{\pi}{2},\label{unstab}
\end{equation}
where $x = (1 - 16\zeta)^{1/4}$.
Having calculated the stability correction $\Psi_m(\zeta)$, we update the estimate for the friction velocity \cite{stull2012introduction}
\begin{equation}
u_*=\frac{k\cdot u}{\ln(z_u/z_{0m})-\Psi_m(\zeta)}.\label{newu}
\end{equation}
This new $u_*$ value is then fed back into Eq. (\ref{lob}) for the Obukhov length $L_{\mathrm{Ob}}$, generating a new $\zeta$, a new $\Psi_m(\zeta)$, and a subsequent new $u_*$. 
This continuous feedback loop iteratively converges on the true values of $u_*$ and $L_{\mathrm{Ob}}$.

\subsubsection{Optical turbulence mapping and refractive index derivation}
Finally, we project the resolved dynamic and thermodynamic structure of the boundary layer onto the atmosphere's optical properties. 
Within the inertial subrange of the turbulence spectrum, the scale at which energy cascades from larger anisotropic eddies to smaller isotropic eddies without viscous dissipation, the spatial structure function of temperature adheres to Kolmogorov's celebrated $2/3$ power law \cite{tatarskii1971effects}.
The amplitude of these rapid temperature fluctuations is quantified by the temperature structure parameter $C_T^2$ \cite{andreas1988estimating}
\begin{equation}
C_T^2=T_*^2\cdot z^{-2/3}\cdot g_T(\zeta), \label{ct}
\end{equation}
the temperature scaling parameter $T_*$ is given by \cite{stull2012introduction}
\begin{equation}
T_*=-\frac{H}{\rho\cdot C_p\cdot u_*}, \label{tstar}
\end{equation}
and the similarity functions $g_T(\zeta)$ is given by \cite{andreas1988estimating}
\begin{equation}
g_T(\zeta)=
\begin{cases}
4.9(1-6.1\zeta)^{-2/3} & \mathrm{for}\quad\zeta\leq0 \\
4.9(1+2.2\zeta^{2/3}) & \mathrm{for}\quad\zeta\geq0
\end{cases}.\label{simt}
\end{equation}

However, in the stable nocturnal boundary layer, the standard MOST encounters a physical singularity under extremely calm conditions $u_* \to 0$ \cite{mahrt1999stratified,grachev2005stable}. 
Physically, the macroscopic sensible heat flux $H$ during night-time is predominantly driven by radiative cooling and molecular conduction rather than mechanical turbulent eddies. 
Substituting the macroscopic $H$ directly into the MOST framework causes $T_*$ to unphysically diverge, which subsequently leads to a severe overestimation of the temperature structure parameter $C_T^2$. 
To resolve this divergence, the mechanically driven turbulent exchange must be isolated \cite{mahrt1999stratified}. 
Under stable stratification, turbulent mixing relies on mechanical shear $u_*$ to overcome negative buoyancy; thus, it is a primary proxy for the transition to intermittent turbulence \cite{van2012cessation}. 
Consequently, we apply a phenomenological Gaussian damping function to extract the turbulent fraction of the sensible heat flux
\begin{equation}
H_{turbulent}=H\left\lbrace1-\exp\left[-\left(\frac{u_*}{u_{scale}}\right)^2\right]\right\rbrace,\label{hten}
\end{equation}
where $u_{\mathrm{scale}} = 0.1 \, \mathrm{m/s}$ is an empirical threshold for intermittency \cite{sun2012turbulence}.
This formulation ensures $T_* \propto u_*$ as $u_* \to 0$, smoothly regularizing the MOST singularity. 

The final step in the derivation converts the thermal turbulence intensity, quantified by the temperature structure parameter $C_T^2$, into the optical turbulence intensity, quantified by the refractive index structure constant  $C_n^2$ \cite{roddier1981v,andrews2005laser}
\begin{equation}
C_n^2=\left(\frac{79\times10^{-6}\cdot P_{hPa}}{(T+273.15)^2}\right)^2\cdot C_T^2,\label{cn2cnt}
\end{equation}
where $P_{hPa}=P/100$.

\subsection{Deep learning for spatial generalization across diverse climates}
\subsubsection{Network architecture}
\begin{figure}
    \centering
    \includegraphics[width=0.7\columnwidth]{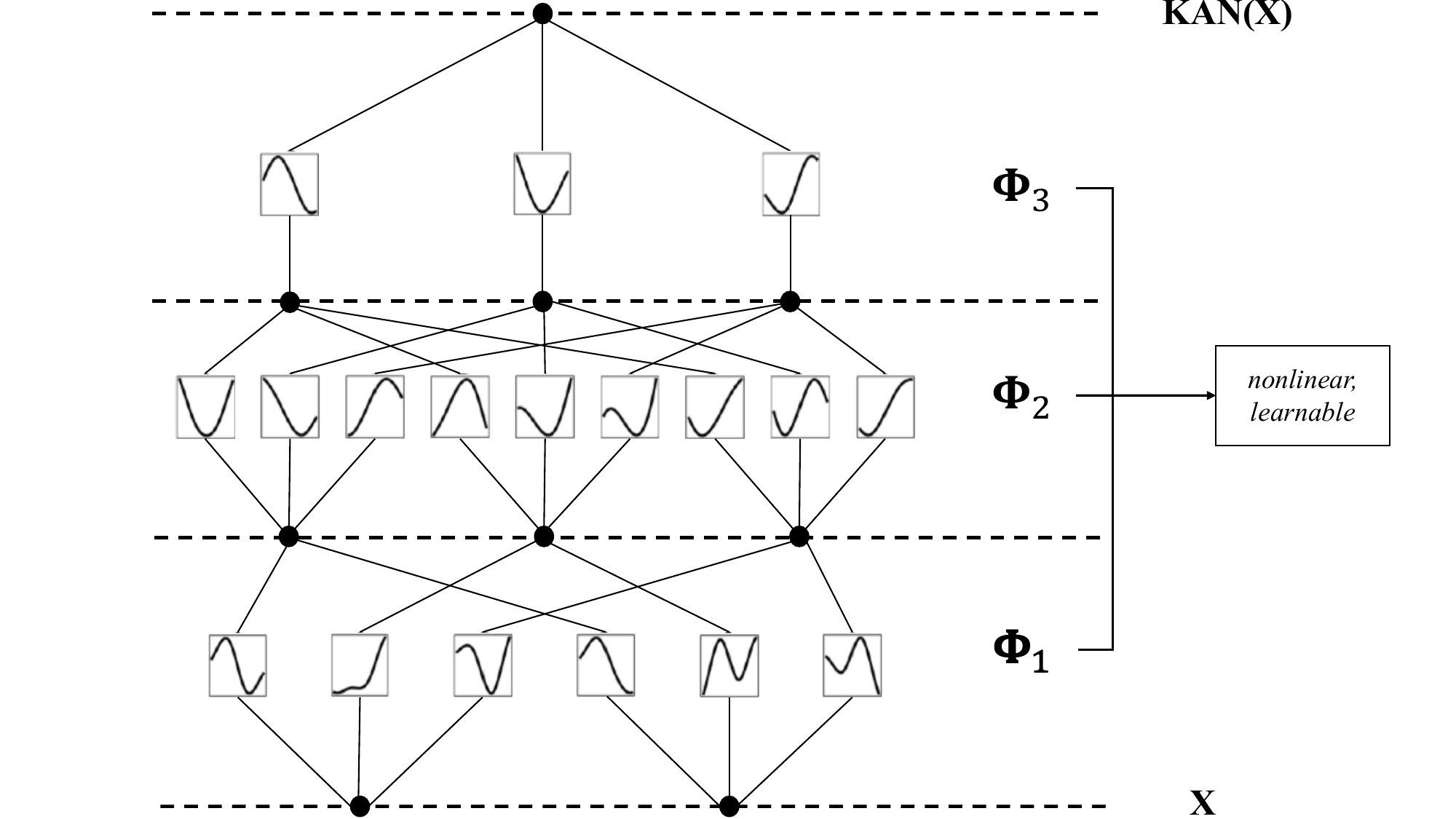}  
    \caption{Architecture of the KAN.}
    \label{fig:KAN}
\end{figure}

In this section, we provide the detailed architecture of the KAN model 
and parameters of our network.
The overall architecture of KAN is illustrated in Fig.~\ref{fig:KAN},
Its design philosophy differs fundamentally from that of MLP: MLP performs information transformation through linear weight matrices between layers and applies fixed nonlinear activation functions (e.g., ReLU) at each neuron (node). 
The core innovation of KAN is the complete elimination of linear weight matrices—each scalar weight in the conventional sense is replaced by a learnable univariate function, and these activation functions are placed on the edges of the network, while nodes perform only summation. 

Specifically, an $L$-layer KAN consists of a set of nodes and the edges connecting them, with the network shape described by the array $[n_0, n_1, \ldots, n_L]$: $n_0$ corresponds to the input dimension, which equals 6 for the meteorological parameters in this task; 
$n_1, n_2, \ldots, n_{L-1}$ correspond to the hidden layer dimensions, determining the model's capacity to learn higher-order interactions and nonlinear mappings among the input variables; in this study, a single hidden layer with $n_1 = 10$ is employed;
$n_L$ corresponds to the output dimension, which equals 1, representing the $\log C_n^2$ value. 
$x_{l,i}$ denotes the value of the $i$-th node in layer $l$, and the activation function on the edge connecting node $(l, i)$ and node $(l+1, j)$ is denoted as $\phi_{l,j,i}$. Then the value of the $j$-th node in layer $l+1$ is computed as: 
\begin{equation}
    x_{l+1,j} = \sum_{i=1}^{n_l} \phi_{l,j,i}(x_{l,i}), \quad j = 1, \ldots, n_{l+1}.
    \label{eq:kan_forward}
\end{equation}
The output of the entire KAN can be expressed as the composition of these function layers:
\begin{equation}
    \mathrm{KAN}(\mathbf{x}) = \left(\Phi_{L-1} \circ \Phi_{L-2} \circ \cdots \circ \Phi_0\right)(\mathbf{x}),
    \label{eq:kan_composition}
\end{equation}
where $\Phi_l$ is the function matrix composed of $\{\phi_{l,j,i}\}$, which entirely replaces the alternating structure of linear transformations and fixed nonlinearities in MLP.
Each edge activation function $\phi(x)$ adopts a learnable parametric form. The B-spline parameterization used in this study is given by:
\begin{equation}
    \phi(x) = w_b \cdot b(x) + w_s \cdot \mathrm{spline}(x),
    \label{eq:bspline_param}
\end{equation}
where $b(x)$ is a base function (e.g., SiLU) serving as a residual connection to ensure global behavior, $\mathrm{spline}(x)$ is a linear combination of B-spline basis functions defined on an adjustable grid with learnable coefficients, and $w_b$ and $w_s$ are learnable scaling factors. 
More explicitly, the B-spline component can be expanded as:
\begin{equation}
    \mathrm{spline}(x) = \sum_{i} c_i B_i(x),
    \label{eq:spline_expansion}
\end{equation}
where $B_i(x)$ are the B-spline basis functions and $c_i$ are the trainable coefficients, enabling high-precision nonlinear approximation through local spline combinations.
The locality of splines enables the model to finely capture local features, and the grid can be dynamically extended during training, thereby enhancing the model's representational capacity without increasing the network depth.

\begin{figure}
\centering
\includegraphics[width=0.9\columnwidth]{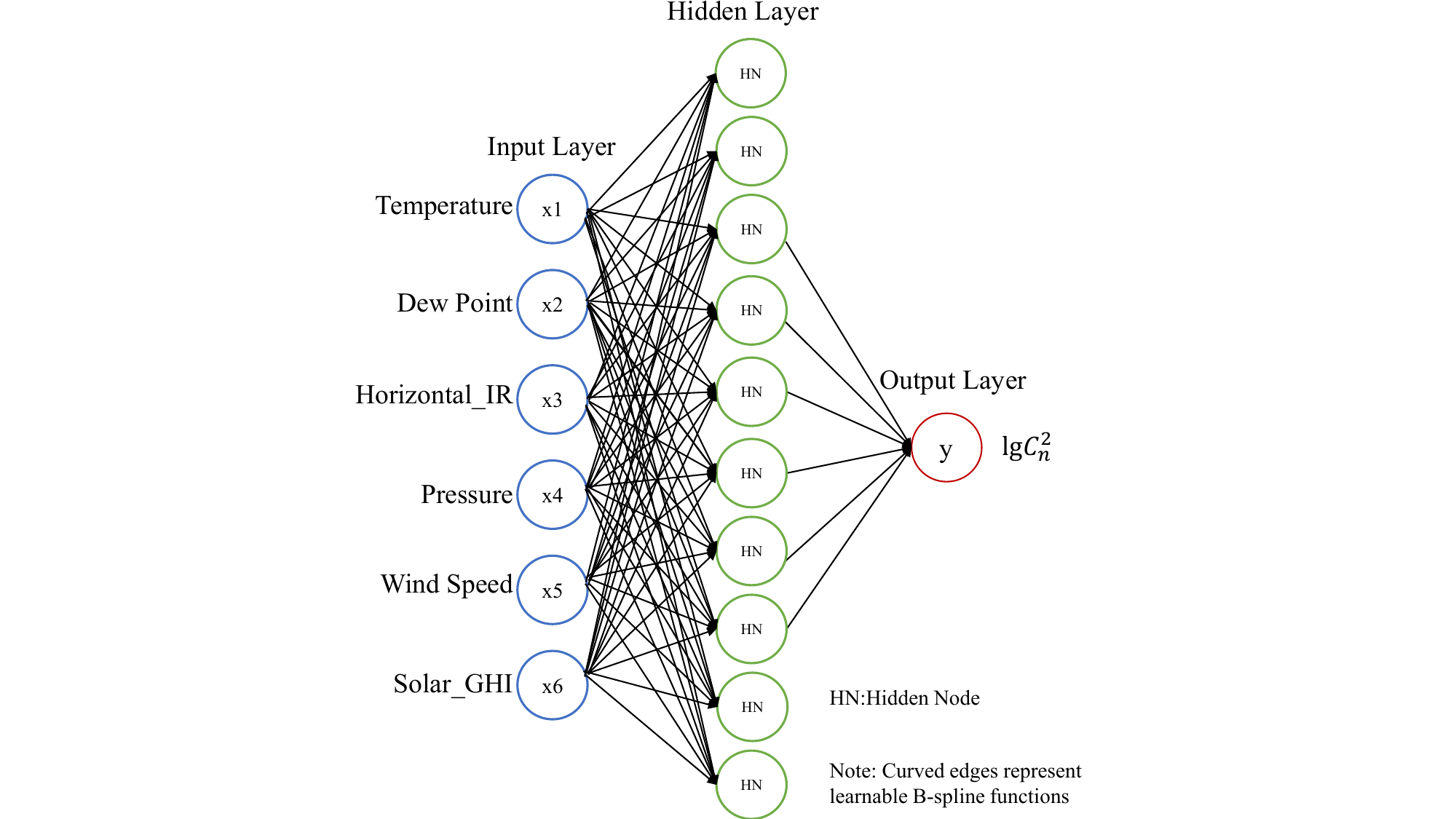}  
\caption{Employed neural network architecture.}\label{fig:Network}
\end{figure}

We construct a KAN model with a specified depth, whose architecture is illustrated in Fig.~\ref{fig:Network}. 
As depicted, the network comprises three types of layers---input, hidden, and output---connected by edges. 
The input layer contains 6 meteorological parameter nodes, corresponding to temperature, dew point, horizontal infrared radiation, atmospheric pressure, wind speed, and global horizontal irradiance, and the hidden layer is set to one layer with 10 nodes, designed to learn the higher-order interactions and nonlinear mapping relationships among meteorological variables. 
The output layer contains a single node corresponding to the logarithmic value of $C_n^2$; adopting the logarithmic form effectively compresses the multi-order-of-magnitude dynamic range typically exhibited by $C_n^2$, thereby stabilizing the training process. 
The edges connecting the nodes do not employ the fixed scalar weights used in conventional neural networks; 
instead, they are composed of learnable nonlinear activation functions parameterized by B-splines. 
Specifically, the B-spline order is set to $k=3$, and the domain of each activation function is uniformly partitioned into 5 intervals.

\subsubsection{Details of experiment}
In this section, we present the formulas for the quantitative experimental evaluation metrics and a comprehensive description of the experimental data, including the specific sites and detailed time information.
\begin{table}[htbp]
\centering
\caption{Temporal distribution of the training set across twelve cities}
\label{tab:training_distribution}
\begin{tabular*}{\columnwidth}{l@{\extracolsep{\fill}}ccc}
\toprule
\textbf{City} & \textbf{Time Coverage (Month:Year)} & \textbf{Span} \\
\midrule
\multirow{2}{*}{Beijing} & Jan 2018, Feb 1979, Mar 1973, Apr 2019, May 1994, Jun 1960 & \multirow{2}{*}{1960--2019} \\
                        & Jul 2011, Aug 2005, Sep 2004, Oct 1984, Nov 1982, Dec 2019 & \\
\midrule
\multirow{2}{*}{Changsha} & Jan 2006, Feb 2011, Mar 2016, Apr 2023, May 2006, Jun 2016 & \multirow{2}{*}{2005--2023} \\
                          & Jul 2008, Aug 2007, Sep 2005, Oct 2012, Nov 2021, Dec 2017 & \\
\midrule
\multirow{2}{*}{Chengdu} & Jan 1988, Feb 1963, Mar 2004, Apr 2020, May 1987, Jun 1961 & \multirow{2}{*}{1961--2020} \\
                         & Jul 1997, Aug 2004, Sep 2012, Oct 1989, Nov 1984, Dec 1992 & \\
\midrule
\multirow{2}{*}{Guangzhou} & Jan 1960, Feb 2011, Mar 2000, Apr 2017, May 2000, Jun 1991 & \multirow{2}{*}{1960--2020} \\
                           & Jul 2012, Aug 2016, Sep 2013, Oct 1986, Nov 2003, Dec 2020 & \\
\midrule
\multirow{2}{*}{Hangzhou} & Jan 1997, Feb 1997, Mar 2009, Apr 1992, May 2006, Jun 1960 & \multirow{2}{*}{1960--2009} \\
                          & Jul 2005, Aug 1996, Sep 2000, Oct 1978, Nov 2003, Dec 1963 & \\
\midrule
\multirow{2}{*}{Harbin} & Jan 2022, Feb 2009, Mar 2020, Apr 2021, May 2004, Jun 2015 & \multirow{2}{*}{2004--2022} \\
                        & Jul 2016, Aug 2007, Sep 2015, Oct 2014, Nov 2013, Dec 2011 & \\
\midrule
\multirow{2}{*}{Wuhan} & Jan 1980, Feb 2001, Mar 1975, Apr 2019, May 1974, Jun 2001 & \multirow{2}{*}{1974--2019} \\
                       & Jul 2003, Aug 1976, Sep 2015, Oct 1976, Nov 1985, Dec 2006 & \\
\midrule
\multirow{2}{*}{Kunming} & Jan 1975, Feb 1975, Mar 1988, Apr 1975, May 1960, Jun 1973 & \multirow{2}{*}{1960--2022} \\
                         & Jul 2014, Aug 1984, Sep 2003, Oct 2022, Nov 1989, Dec 1976 & \\
\midrule
\multirow{2}{*}{Shanghai} & Jan 2007, Feb 2015, Mar 2016, Apr 2006, May 2019, Jun 2007 & \multirow{2}{*}{2006--2020} \\
                          & Jul 2012, Aug 2019, Sep 2009, Oct 2013, Nov 2014, Dec 2020 & \\
\midrule
\multirow{2}{*}{Urumqi} & Jan 1989, Feb 1985, Mar 2007, Apr 2005, May 2004, Jun 2010 & \multirow{2}{*}{1985--2019} \\
                        & Jul 2006, Aug 2010, Sep 1998, Oct 2005, Nov 1990, Dec 2019 & \\
\midrule
\multirow{2}{*}{Xiamen} & Jan 1980, Feb 1976, Mar 2009, Apr 1979, May 1961, Jun 1960 & \multirow{2}{*}{1958--2013} \\
                        & Jul 2000, Aug 1977, Sep 1958, Oct 1980, Nov 2013, Dec 1978 & \\
\midrule
\multirow{2}{*}{Xi'an} & Jan 1997, Feb 1961, Mar 1963, Apr 1987, May 2015, Jun 1959 & \multirow{2}{*}{1959--2015} \\
                       & Jul 2009, Aug 1996, Sep 1989, Oct 2008, Nov 2004, Dec 1963 & \\
\bottomrule
\end{tabular*}
\end{table}

\begin{table}[htbp]
\centering
\caption{Temporal distribution of the test set across three cities}
\label{tab:test_distribution}
\begin{tabular*}{\columnwidth}{l@{\extracolsep{\fill}}ccc}
\toprule
\textbf{City} & \textbf{Time Coverage (Month Year)} & \textbf{Span} \\
\midrule
\multirow{2}{*}{Hohhot} & Jan 1988, Feb 2001, Mar 1995, Apr 1992, May 2000, Jun 1998 & \multirow{2}{*}{1963--2016} \\
                        & Jul 1989, Aug 1963, Sep 2004, Oct 2007, Nov 2016, Dec 1976 & \\
\midrule
\multirow{2}{*}{Lhasa} & Jan 1985, Feb 1987, Mar 1988, Apr 1988, May 1985, Jun 1978 & \multirow{2}{*}{1977--2014} \\
                       & Jul 1980, Aug 2004, Sep 1983, Oct 2014, Nov 1977, Dec 2000 & \\
\midrule
\multirow{2}{*}{Sanya} & Jan 2013, Feb 2020, Mar 2013, Apr 2012, May 2013, Jun 2017 & \multirow{2}{*}{2005--2022} \\
                       & Jul 2006, Aug 2022, Sep 2014, Oct 2017, Nov 2005, Dec 2020 & \\
\bottomrule
\end{tabular*}
\end{table}
The correlation coefficient ($R_{xy}$), mean squared error ($MSE$), mean absolute error ($MAE$), and bias are four core metrics for measuring model prediction performance.
$R_{xy}$ evaluates the linear correlation between the predicted and true values; 
$MSE$ and $MAE$ characterize the magnitude of prediction errors from different sensitivity perspectives; 
and bias indicates whether the model exhibits a systematic tendency to overestimate or underestimate. 
Together, these metrics form a comprehensive and complementary evaluation system for the task of predicting $\lg C_n^2$.
\begin{equation}
R_{xy} = \frac{1}{N-1} \sum_{i=1}^N \left( \frac{X_i - \bar{X}}{\sigma_X} \right) \left( \frac{Y_i - \bar{Y}}{\sigma_Y} \right),
\end{equation}
\begin{equation}
\mathrm{MSE} = \frac{1}{N} \sum_{i=1}^{N} \left( X_i - \hat{X}_i \right)^2,
\end{equation}
\begin{equation}
\mathrm{MAE} = \frac{1}{N} \sum_{i=1}^{N} \left| X_i - \hat{X}_i \right|,
\end{equation}
\begin{equation}
\mathrm{Bias} = \frac{1}{N} \sum_{i=1}^{N} \left( X_i - \hat{X}_i \right),
\end{equation}
where $N$ is the sample size; $X_i$ and $\hat{X}_i$ are the individual sample points indexed with $i$, representing the true and estimated values of $\lg C_n^2$, respectively; $Y_i$ denotes the estimated values; $\bar{X}$ is the sample mean; $\sigma_X$ is the sample standard deviation (similarly for $\bar{Y}$ and $\sigma_Y$).

The training set covers twelve major cities in China (Beijing, Changsha, Chengdu, Guangzhou, Hangzhou, Harbin, Wuhan, Kunming, Shanghai, Urumqi, Xiamen, and Xi'an). 
For each city, twelve sample months (from January to December) are selected, with a temporal resolution of one hour — i.e., approximately 720 hourly records per month, 
depending on the number of days in that month. 
The specific temporal distributions for each training city are summarized in Table~\ref{tab:training_distribution}. 
The test set comprises three additional cities (Hohhot, Lhasa, and Sanya), also with twelve sample months per city and hourly resolution, 
as detailed in Table~\ref{tab:test_distribution}. 
Together, the training and test sets span a wide range of geographical regions and temporal periods, providing a robust foundation for model evaluation.

Overall, the earliest data in the training set dates back to September 1958 (Xiamen), and the latest data is April 2023 (Changsha), covering a time span of approximately 65 years. 
The temporal distributions vary across cities — some are concentrated in the recent two decades (e.g., Shanghai 2006--2020, Harbin 2004--2022), while others include historical observations from earlier years (e.g., Kunming 1960, Xi'an 1959). This long-span sampling design provides a rich and diverse sample foundation for the model to learn the long-term variation patterns of atmospheric parameters, particularly $\lg C_n^2$. 
The test set complements the training set by introducing three additional cities — Hohhot (arid continental, 1963--2016), Lhasa (high-altitude plateau, 1977--2014), and Sanya (tropical maritime, 2005--2022) — which span different geographical regions and climatic conditions. This ensures that the model's generalization capability is rigorously evaluated across diverse climatic zones and time frames.

\subsection{Experimental results}
In this section, we present our experimental results, including quantitative evaluations, scatter plots, and time-series comparisons between ground truth and estimated values, to demonstrate that the KAN model achieves higher accuracy and stronger generalization capability. 
\begin{figure*}[!htbp]
    \centering
    \subfloat[]{
        \label{fig:scatter_aga}   
        \includegraphics[width=0.48\textwidth]{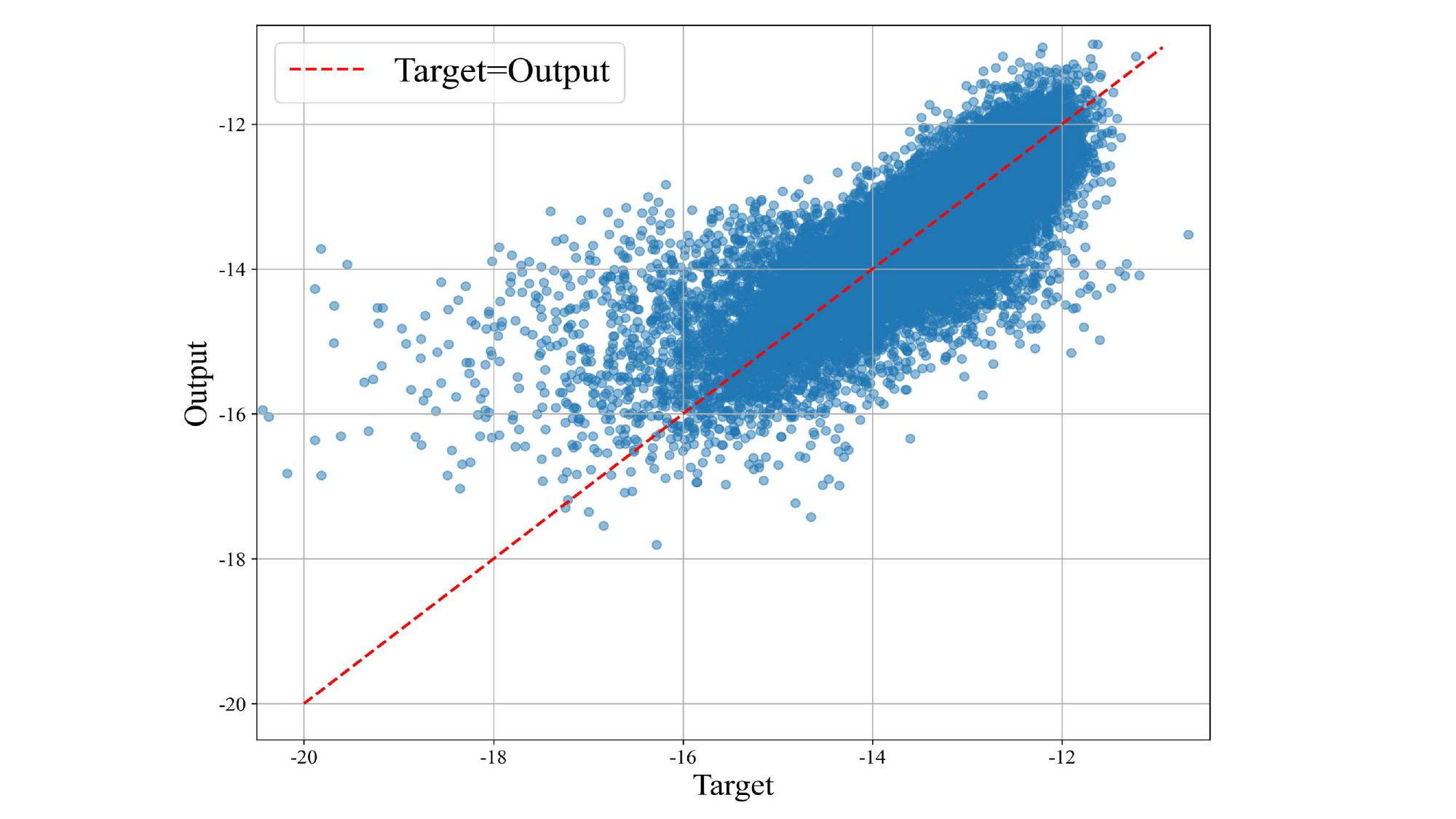}
    } \hfill
    \subfloat[]{
        \label{fig:scatter_kan}
        \includegraphics[width=0.48\textwidth]{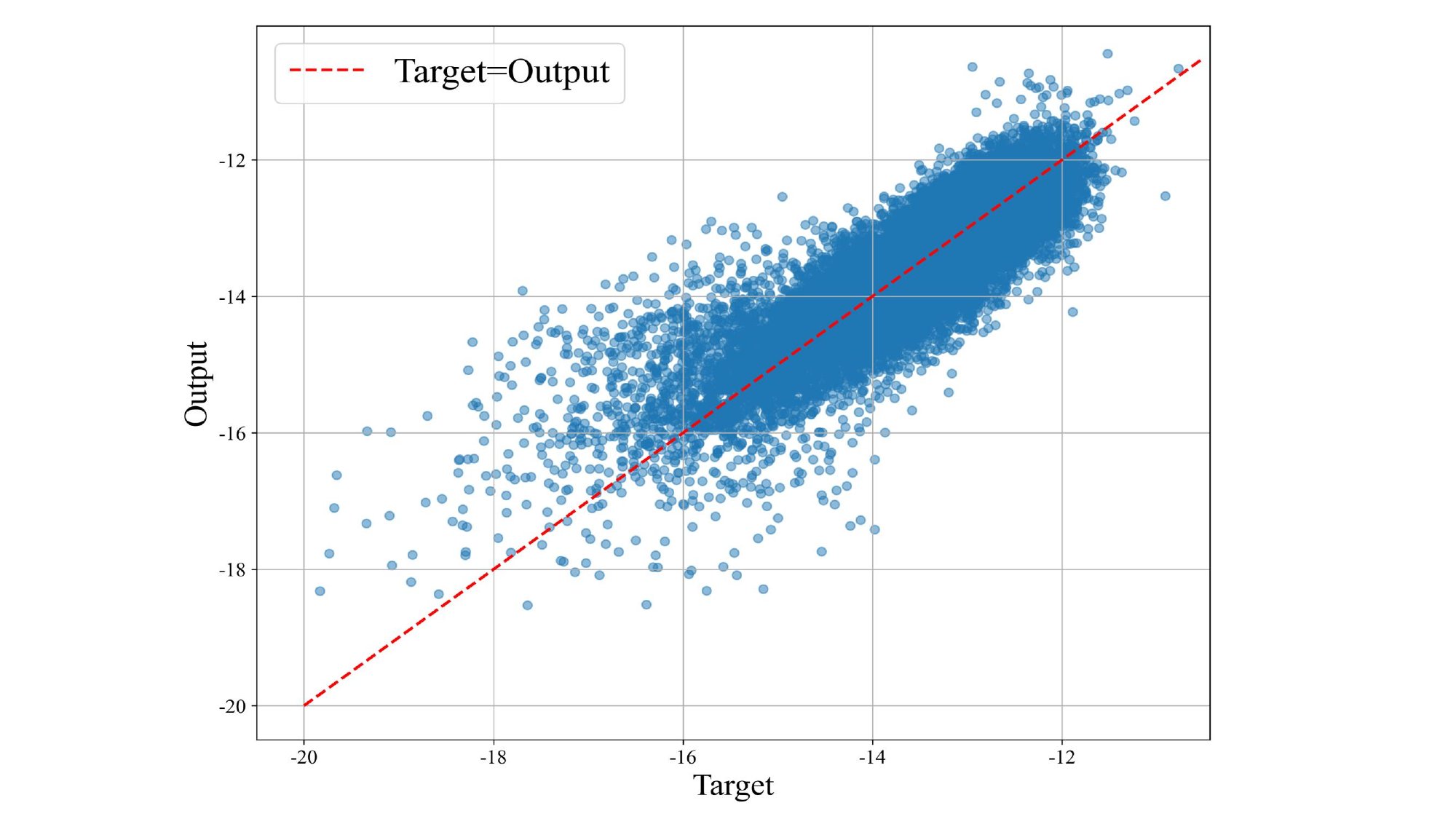}
    }
    \caption{Scatter plot comparison of measured and estimated values for the (a) AGA-BP and (b) KAN models. Both axes are logarithmic, representing the true and estimated values of $\log_{10} C_n^2$,     respectively.}
    \label{fig:Scatter_plot}
\end{figure*}
\begin{figure*}[!htbp]
    \centering
    \includegraphics[width=0.9\textwidth]{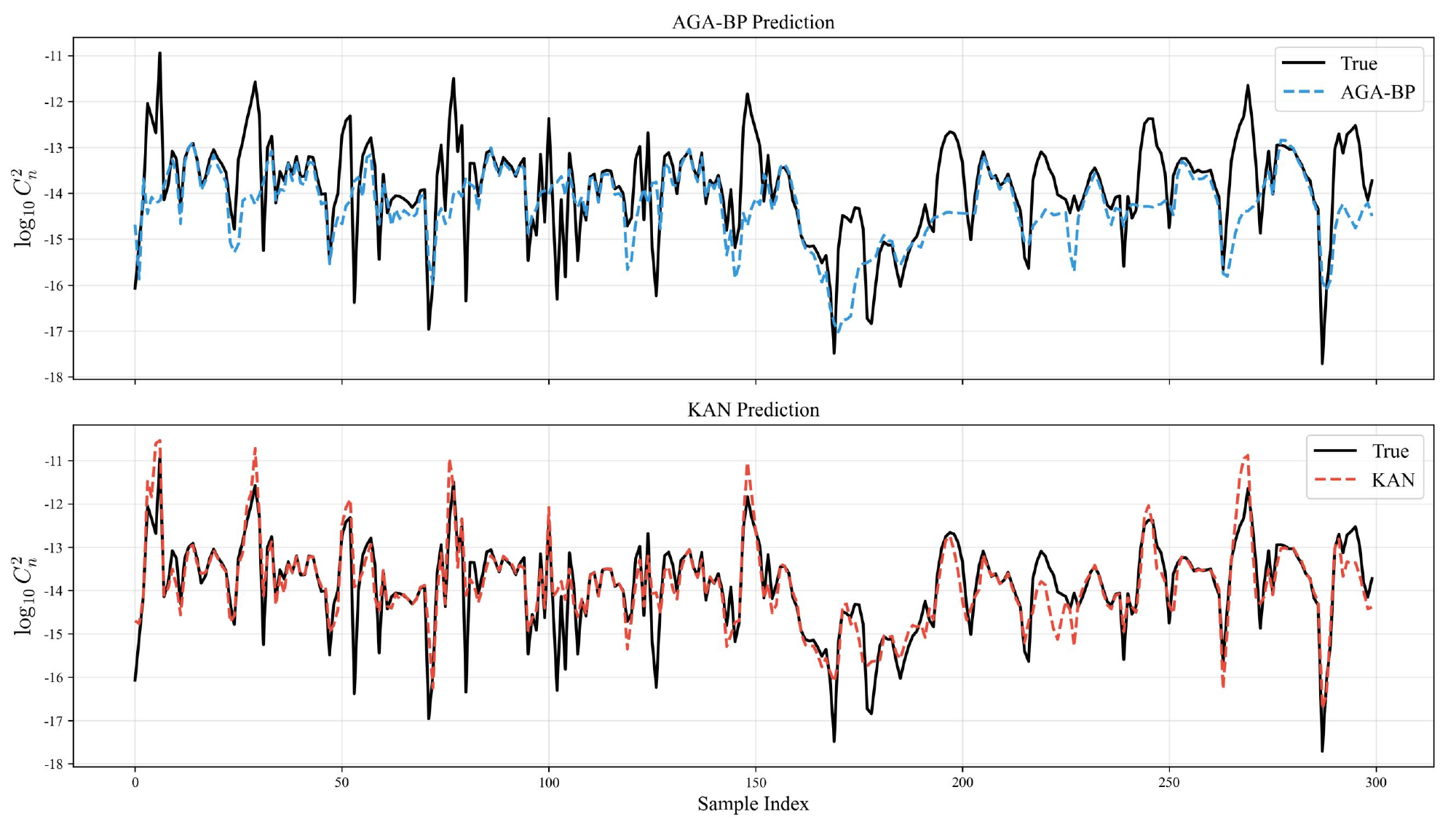}
    \caption{ The horizontal axis represents 300 consecutive samples from the test set, spanning 12 days of $\log C_n^2$ variations, from 09:00 on January 9, 1988 to 22:00 on January 21, 1988, collected from the Hohhot site. The black curve denotes the true values, the blue curve represents the estimates of the AGA-BP model, and the red curve corresponds to the estimates of the KAN model.}
    \label{fig:curve}
\end{figure*}
The conventional AGA-BP is based on MLP architecture \cite{wang2016using,su2020adaptive}.
Fig.~\ref{fig:Scatter_plot} shows the scatter plots of the estimated results of the two models against the true target values on the test set.  
Comparing the two subplots, it can be observed that the scatter points in Fig.~\ref{fig:scatter_kan} are more concentrated around the ideal reference line, exhibiting lower dispersion, fewer outliers, and better alignment with the estimated versus true values. 
This indicates that the corresponding model achieves superior generalization performance on the test set, along with enhanced estimation stability.
As shown in the time series plot Fig.~\ref{fig:curve}, in areas where the changes in $C_n^2$ are relatively gradual, both models have high accuracy. 
However, in areas with sharp fluctuations, the KAN model can more accurately depict the trend of the changes, 
whereas the AGA-BP model has lower accuracy. 
Especially in multiple sections with abrupt increases, the estimated values of the AGA-BP model are significantly lower than the actual values. 
This indicates that the AGA-BP model struggles to capture sudden changes in $C_n^2$, while the KAN model has stronger nonlinear mapping capabilities and greater sensitivity to extreme fluctuations.
\begin{table}[!htbp]
\centering
\renewcommand{\arraystretch}{1.3}
\caption{Quantitative Evaluation between Different Models.}
\label{tab:performance}
\begin{tabular*}{\columnwidth}{l@{\extracolsep{\fill}}cccc}
\toprule
Model & $R_{xy}$ & MSE & Bias & MAE \\
\midrule
AGA-BP & 0.842 & 0.283 & 0.002 & 0.319 \\
KAN & 0.911 & 0.166 & -0.001 & 0.213 \\
\bottomrule
\end{tabular*}
\end{table}

We quantitatively evaluated the reliability of the model estimates using four statistical indicators.
The MSE and MAE are absolute error measures, with lower values indicating better performance. 
The $R_{xy}$ value, closer to 1, indicates a higher degree of linear correlation between the predicted and true values. 
Moreover, a Bias value closer to 0 suggests a smaller systematic error of the model.
Table~\ref{tab:performance} summarizes the calculated values of these indicators for both models.
\subsection{The atmospheric transmittance analysis}
In this section, we outline the application of the elliptic-beam approximation to evaluate the mean error-probability, expressed as $\bar{P}_{\mathrm{err}}=\int_{0}^{\tau_{\mathrm{max}}}d\tau \mathcal{P}(\tau)P_{\mathrm{err}}(\tau)$. 
Comprehensive details regarding this model are provided in Ref. \cite{Vasylyev:2016xmk}. 
To this end, we generate $N$ independent Gaussian random vectors $\mathbf{v}_{i}=(x_{0;i}, y_{0;i}, \Theta_{1;i}, \Theta_{2;i})^{\mathrm{T}}$ and uniformly distributed random angles $\phi_{i}\in[0,\pi/2]$ for $i=1,\ldots,N$. 
Here, $x_{0;i}$ and $y_{0;i}$ denote the random coordinates of the beam centroid, while $\Theta_{1;i}$ and $\Theta_{2;i}$ parameterize the semi-axes $W_{1;i}$ and $W_{2;i}$ of the random ellipses. 
These ellipses model the beam profile after propagation through the atmosphere, governed by the relation $W_{1/2;i}^2=W_0^2\exp(\Theta_{1/2;i})$, where $W_0$ represents the initial beam-spot radius at the transmitter.

Table \ref{tabc} presents the mean values and covariance matrix elements of the random vector $\mathbf{v}_{i}$ for horizontal links, parameterized by the initial beam-spot radius $W_0$, the Fresnel parameter $\Omega=kW_{0}^{2}/2L$, and the Rytov parameter $\sigma_R^2=1.23C_n^2k^{\frac{7}{6}}L^{\frac{11}{6}}$. 
Here, $k$ denotes the optical wavenumber, $L$ represents the propagation distance, and $\gamma=(1+\Omega^2)/\Omega^2$.
\begin{table}[h!]
\caption{The non-zero elements of the covariance matrix and the means}
\renewcommand{\arraystretch}{1.5}
\begin{tabular*}{\textwidth}{@{\extracolsep{\fill}}ccc}
\toprule
& Weak turbulence & Strong turbulence \\
\midrule
$\left\langle \Theta_{1/2;i}\right\rangle$ & 
$\ln\Biggl[\frac{\left(1+2.96 \sigma_R^2\Omega^{\frac{5}{6}}\right)^2}{\Omega^2\sqrt{\left(1+2.96 \sigma_R^2\Omega^{\frac{5}{6}}\right)^2+1.2\sigma_R^2\Omega^{\frac{5}{6}}}}\Biggr]$ & 
$\ln\Biggl[\frac{\bigl(\gamma+1.71 \sigma_R^{\frac{12}{5}}\Omega^{{-}1}{-}2.99 \sigma_R^{\frac{8}{5}}\Omega^{{-}1}\bigr)^2}{\sqrt{\bigl(\gamma+1.71 \sigma_R^{\frac{12}{5}}\Omega^{{-}1}{-}2.99 \sigma_R^{\frac{8}{5}}\Omega^{{-}1}\bigr)^2+3.24 \gamma\sigma_R^{\frac{12}{5}}\Omega^{-1}}}\Biggr]$\\

$\left\langle\Delta x_{0;i}^2\right\rangle,\left\langle\Delta y_{0;i}^2\right\rangle$ & 
$0.33\,W_0^2 \sigma_R^2 \Omega^{-\frac{7}{6}}$ & 
$0.75\,W_0^2 \sigma_R^{\frac{8}{5}} \Omega^{-1}$ \\

$\left\langle \Delta \Theta_{1/2;i}^2\right\rangle$ & 
$\ln\Biggl[1+\frac{1.2\sigma_R^2\Omega^{\frac{5}{6}}}{\left(1+2.96\sigma_R^2\Omega^{\frac{5}{6}}\right)^2}\Biggr]$ & 
$\ln\Biggl[1+\frac{13.14 \gamma\sigma_R^{12/5}\Omega^{-1}}{\bigl(\gamma+1.71 \sigma_R^{\frac{12}{5}}\Omega^{{-}1}{-}2.99 \sigma_R^{\frac{8}{5}}\Omega^{{-}1}\bigr)^2}\Biggr]$ \\

$\left\langle \Delta \Theta_{1;i}\Delta \Theta_{2;i}\right\rangle$ & 
$\ln\Biggl[1-\frac{0.8\sigma_R^2\Omega^{\frac{5}{6}}}{\left(1+2.96\sigma_R^2\Omega^{\frac{5}{6}}\right)^2}\Biggr]$ & 
$\ln\Biggl[1+\frac{0.65\gamma \sigma_R^{12/5}\Omega^{-1}}{\bigl(\gamma+1.71 \sigma_R^{\frac{12}{5}}\Omega^{{-}1}{-}2.99 \sigma_R^{\frac{8}{5}}\Omega^{{-}1}\bigr)^2}\Biggr]$ \\
\bottomrule
\end{tabular*}
\label{tabc}
\end{table}
Specifically, the table delineates the theoretical results for both weak and strong turbulence regimes. 
The weak turbulence approximations $\sigma_{R}^{2}\lesssim1$ apply to short propagation distances; in near-surface optical communication scenarios, these conditions are typically met during nighttime operations. 
Conversely, the strong turbulence results $\sigma_{R}^{2}\gg1$ govern long-distance propagation, which, for near-surface links, corresponds to daytime operations under clear-sky conditions.
\begin{figure}[htbp]
\centering
\includegraphics[width=0.55\columnwidth]{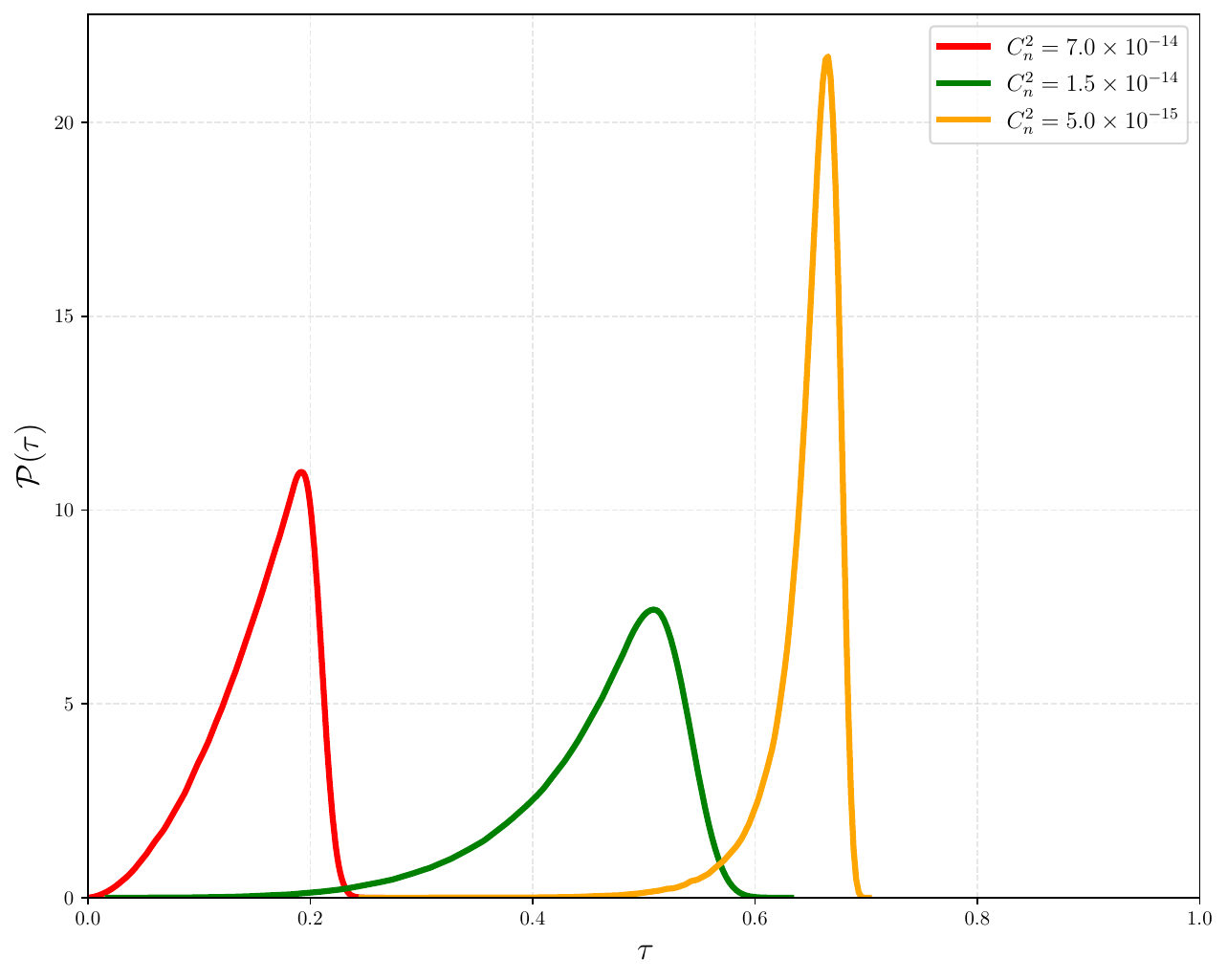}  
\caption{Probability distribution of the atmospheric transmission given by the elliptic beam model for different $C_n^2$
.}\label{pdtf}
\end{figure}

Based on the generated sampling data, the mean error-probability 
is approximated as 
\begin{equation}
\bar{P}_{\mathrm{err}}=\frac{1}{N}\sum_{i=1}^{N}{P}_{\mathrm{err}}\left[\tau(\mathbf{v}_{i},\phi_i)\right].\label{perr}
\end{equation}
The transmittance function $\tau(\mathbf{v},\phi)$ is given by
\begin{equation}
\begin{aligned}\tau(\mathbf{v},\phi)&=\tau_{0}(\Theta_{1},\Theta_{2})\times\exp\left\{-\left[\frac{r_{0}/a}{R\left(\frac{2}{W_{\mathrm{eff}}(\Theta_{1},\Theta_{2},\phi)}\right)}\right]^{\lambda\left(\frac{2}{W_{\mathrm{eff}}(\Theta_{1},\Theta_{2},\phi)}\right)}\right\}
\end{aligned},\label{tautau}
\end{equation}
where $r_0=\sqrt{x_0^2+y_0^2}$ is the distance between beam and aperture centers, $a$ is the radius of the receiver aperture.
The further parameters introduced in this function are given by
\begin{equation}W_{\mathrm{eff}}^{2}(\Theta_{1},\Theta_{2},\phi)=4a^{2}\left[\mathcal{W}\left(\frac{4a^{2}}{W_{1}(\Theta_{1})W_{2}(\Theta_{2})}\right)\times\mathrm{e}^{\frac{a^{2}}{W_{1}^{2}(\Theta_{1})}\{1+2\cos^{2}\phi\}}\mathrm{e}^{\frac{a^{2}}{W_{2}^{2}(\Theta_{2})}\{1+2\sin^{2}\phi\}})\right]^{-1},\label{weff}
\end{equation}
\begin{align}\tau_0(\Theta_1,\Theta_2)=&1-\mathrm{I}_{0}\left(a^{2}\left[\frac{1}{W_{1}^{2}(\Theta_{1})}-\frac{1}{W_{2}^{2}(\Theta_{2})}\right]\right)\mathrm{e}^{-a^{2}\left[\frac{1}{W_{1}^{2}(\Theta_{1})}+\frac{1}{W_{2}^{2}(\Theta_{2})}\right]}-2\left[1-\mathrm{e}^{-\frac{a^2}{2}\left(\frac{1}{W_1(\Theta_1)}-\frac{1}{W_2(\Theta_2)}\right)^2}\right]\\ \notag
&\times\exp\left\{-\left[\frac{\frac{(W_1(\Theta_1)+W_2(\Theta_2))^2}{\mid W_1^2(\Theta_1)-W_2^2(\Theta_2)\mid}}{R\left(\frac{1}{W_1(\Theta_1)}-\frac{1}{W_2(\Theta_2)}\right)}\right]^{\lambda\left(\frac{1}{W_1(\Theta_1)}-\frac{1}{W_2(\Theta_2)}\right)}\right\},\label{maxtau}
\end{align}
\begin{equation}
R(\xi)=\left[\ln\left(2\frac{1-\exp[-\frac{1}{2}a^{2}\xi^{2}]}{1-\exp[-a^{2}\xi^{2}]\mathrm{I}_{0}(a^{2}\xi^{2})}\right)\right]^{-\frac{1}{\lambda(\xi)}},\label{funcr}
\end{equation}
\begin{equation}
\begin{aligned}\lambda(\xi)=2a^{2}\xi^{2}\frac{\mathrm{e}^{-a^{2}\xi^{2}}\mathrm{I}_{1}(a^{2}\xi^{2})}{1-\exp[-a^{2}\xi^{2}]\mathrm{I}_{0}(a^{2}\xi^{2})}\times\left[\ln\left(2\frac{1-\exp[-\frac{1}{2}a^{2}\xi^{2}]}{1-\exp[-a^{2}\xi^{2}]\mathrm{I}_{0}(a^{2}\xi^{2})}\right)\right]^{-1}
\end{aligned}
\end{equation}
Here, $\mathcal{W}(\xi)$ is the Lambert $W$ function \cite{corless1996lambert}, and $I_{i}(\xi)$ is the modified Bessel function of $i$th order.

As illustrated in Fig. \ref{pdtf}, the presence of PDTs is evident for three distinct turbulence strengths, characterised by the $C_n^2$. 
The corresponding distributions exhibit average transmission coefficients $\langle\tau\rangle$ of 0.16, 0.46, and 0.65, respectively. 
The parameter values used in the channel are shown in Table \ref{Tabph}.

\begin{table}[htbp]
\centering
\caption{Physical parameters of atmospheric QI channel}
\label{Tabph}
\renewcommand{\arraystretch}{1.2}
\begin{tabular*}{\linewidth}{@{\extracolsep{\fill}} ccc @{}}
\toprule
Parameter &  Description  & Value \\
\midrule
$z$ &  Channel height & $1.5\,\mathrm{m}$ \\
$\lambda$ & Wavelength &  $809\,\mathrm{nm}$ \\
$L$ & Propagation distance &  $1.6\,\mathrm{km}$ \\
$W_0$ & Initial spot radius &  $20\,\mathrm{mm}$ \\
$a$ & Aperture radius &  $40\,\mathrm{mm}$ \\
$N_S$ & Mean signal photon number &  $0.01$ \\
$N_B$ & Mean thermal photon number &  $20$ \\
$\Lambda$ & Effective mode number &  $10^7$ \\
\bottomrule
\end{tabular*}
\end{table}

\end{document}